\documentclass[12pt,fleqn]{article}

\usepackage{amsmath, amssymb,graphicx}

  \newcommand{\s}{\sigma}

\def\V{{\cal V}}
  \newcommand{\N}{\mathcal{N}}

\def\D{{\cal D}}

\newcommand{\CC}{\mathbb{C}}
\newcommand{\RR}{\mathbb{R}}
\newcommand{\ZZ}{\mathbb{Z}}

\begin{document}


\thispagestyle{empty}
\renewcommand{\thefootnote}{\fnsymbol{footnote}}

{\hfill \parbox{3cm}{
        HU-EP-02/57 \\
        MIT-CTP-3333 \\
}}

\bigskip\bigskip\bigskip

\begin{center} \noindent \Large \bf

{\bf  (De)constructing Intersecting
M5-Branes}
\end{center}

\bigskip\bigskip\bigskip\bigskip

\centerline{ \normalsize \bf Neil R. Constable$\,^{a}$, Johanna
Erdmenger$\,^b$,
Zachary Guralnik$\,^{b}$ and
Ingo Kirsch$\,^{b}$\footnote[1]{\noindent \tt
constabl@lns.mit.edu, jke@physik.hu-berlin.de,
zack@physik.hu-berlin.de,  ik@physik.hu-berlin.de} }

\bigskip
\bigskip\bigskip

\bigskip\bigskip

\centerline{$^a$ \it
Center for Theoretical Physics and Laboratory for Nuclear Science    }
\centerline{ \it Massachusetts Institute of Technology}
\centerline{\it 77 Massachusetts Avenue}
\centerline{ \it Cambridge, {\rm MA}  02139, USA}
\bigskip
\centerline{$^b$ \it Institut f\"ur Physik}
\centerline{\it Humboldt-Universit\"at zu Berlin}
\centerline{\it Invalidenstra{\ss}e 110}
\centerline{\it D-10115 Berlin, Germany}
\bigskip\bigskip

\bigskip\bigskip

\renewcommand{\thefootnote}{\arabic{footnote}}

\centerline{\bf \small Abstract}
\medskip

{\small We describe intersecting M5-branes, as well as M5-branes
wrapping the holomorphic curve $xy=c$, in terms of a limit of a
defect conformal field theory with two-dimensional $(4,0)$
supersymmetry.  This defect CFT describes the low-energy theory of
intersecting D3-branes at a $\CC^2/\ZZ_k$ orbifold.  In an
appropriate $k\rightarrow \infty$ limit, two compact spatial
directions are generated. By identifying moduli of the M5-M5
intersection in terms of those of the defect CFT, we argue that the
$SU(2)_L$ R-symmetry of the $(4,0)$ defect CFT matches the $SU(2)$
R-symmetry of the ${\cal N} =2, d=4$ theory of the M5-M5
intersection.  We find a 't Hooft anomaly in the $SU(2)_L$
R-symmetry, suggesting that tensionless strings give rise to an
anomaly in the $SU(2)$ R-symmetry of intersecting M5-branes.}

\newpage


\newpage
\section{Introduction}
\setcounter{equation}{0}

In recent years string theory has suggested the existence of novel
interacting conformal theories in diverse dimensions. In many
cases, a Lagrangian description of these theories is lacking. A
particularily interesting example is the six-dimensional theory with
$(2,0)$ supersymmetry describing the low energy limit of IIB
string theory on an $A_n$ singularity \cite{Witten20}, as well as
the decoupling limit of multiple parallel M5-branes
\cite{Strominger}.  Although this theory is believed to be a local
quantum field theory, obstructions to finding a Lagrangian
description arise because of difficulties in constructing a
non-abelian generalization of a chiral two-form (see for example
\cite{BHS}). The spectrum includes tensionless BPS strings, which
are in some sense the ``off-diagonal'' excitations of the
non-abelian chiral two-form. Until recently, the only known
formulation of theory was in terms of a M(atrix) model describing
its discrete light cone quantization \cite{Aharony}. More
recently, an alternative formulation was found \cite{Arkani-Hamed,CET}
using a procedure known as (de)construction \cite{Arkani-Hamed1,Csaki1}.
In this
approach, the $(2,0)$ theory is obtained as a limit of ${\cal N}
=2, d=4$ superconformal Yang-Mills theories described by a
circular quiver diagram. This limit entails taking the number $k$
of nodes in the diagram to infinity, while scaling the gauge
coupling as $\sqrt{k}$.  At the same time one goes increasingly
far out onto the Higgs branch, on which the gauge group is broken
from $SU(N)^k$ to the diagonal $SU(N)$. The quiver diagram can
then be viewed a discretization of an extra spatial circle, which
is believed to become continuous as $k\rightarrow\infty$. The S-duality of
the
theory implies the generation of yet another discretized circular
dimension which also becomes continuous as $k\rightarrow \infty$.

Even more poorly understood than the $(2,0)$ theory is the one
which describes the low energy dynamics of intersecting M5-branes.
In addition to a non-abelian chiral two-form,  this theory has
tensionless strings localized at the intersection corresponding to
M2-branes stretched between the M5-branes \cite{Hanany}.  These
tensionless strings are in some sense fundamental,  as they are
not excitations of a chiral two-form.  The only known formulation
of the M5-M5 intersection is the DLCQ M(atrix) description
proposed in \cite{Kachru}.

Here we shall present the (de)construction of the M5-M5
intersection,  which is a natural extension of the
(de)construction of parallel M5-branes discussed in
\cite{Arkani-Hamed,CET}.  This will be accomplished by taking a
$k \rightarrow \infty$ limit of the theory describing intersecting
D3-branes at a $\CC^2/\ZZ_k$ orbifold. At a certain point in the
moduli space, two compact latticized extra dimensions are
generated. In an appropriate $k\rightarrow \infty$ limit, we expect that
the extra directions become continuous, such that the intersection of
four-dimensional world volumes over $1+1$ dimensions becomes an
intersection of six-dimensional world volumes over $1+3$
dimensions.

The infrared dynamics of the D3-D3 intersection at a $\CC^2/\ZZ_k$
orbifold is described by a defect conformal field theory with
two-dimensional $(4,0)$ supersymmetry.  This theory belongs to an
interesting class of conformal field theories with defects which
have recently been studied in a variety of contexts
\cite{Sethi,GanorSethi,KapustinSethi,Karchrandall1,Karchrandall2,
Karchrandall3,DeWolfe,EGK,Lee,Skenderis,Bachas,Quella,Mateos,CEGK}.
The action of this $(4,0)$ theory is readily constructed in $(2, 0)$
superspace, starting from the action for the D3-D3 intersection in
flat space which was constructed in \cite{CEGK}.
The field content of the $(4,0)$ theory is summarized by a quiver
(or ``moose'') diagram consisting of two concentric rings, and
spokes stretching between the inner and outer rings. For large $k$
this gives rise to a discretized version of the field theory
corresponding to the low-energy limit of the M5-M5 intersection.
The spokes in the quiver diagram
will be seen to correspond to strings localized at the M5-brane
intersection.

Moreover, we examine the relation between the moduli space of
vacua of the $(4,0)$ defect conformal field theory and that of the
M5-M5 intersection. On a particular part of the Higgs branch of
the defect CFT, the resolution of the intersection to a
holomorphic curve $xy =c$ can be seen very explicitly from
F-flatness conditions.  This point in the Higgs branch corresponds
to a vacuum of the M5-M5 theory in which tensionless strings have
condensed. By going to another point on the Higgs branch of the
defect CFT for which the string tension in the M5-M5 theory is
non-zero, we will be able to match the $SU(2)_L$ R-symmetry of the
$(4,0)$ theory with the $SU(2)$ R-symmetry of the M5-M5
intersection, which has ${\cal N} =2, d=4$ supersymmetry.

The chiral nature of the theory which deconstructs the M5-M5
intersection is a bit surprising,  and should have physical
consequences.  Using the $(4,0)$ dCFT,  we will search for 't
Hooft anomalies in the R-symmetry of the M5-M5 intersection. One
reason to be interested in R-symmetry anomalies is that they are
related by supersymmetry to the Weyl anomaly and to black hole
entropy \cite{FHMM,HMM}.  Their existence also influences the low
energy effective theory at certain points in the moduli space
through Wess-Zumino terms which appear upon integrating out
degrees of freedom responsible for the anomaly
\cite{intriligator}.  It turns out there is a 't Hooft anomaly in
the $SU(2)_L$ R-symmetry of the $(4,0)$ dCFT, under which only
left handed two-dimensional fermions are charged. Assuming a
finite continuum limit, this anomaly should be interpreted as an
$SU(2)$ R-symmetry anomaly due to tensionless strings in four
dimensions. Although there are no $SU(2)$ anomalies in local
quantum field theories in four dimensions,  the possibility is not
excluded for a four-dimensional theory of tensionless strings.
Unfortunately we can not yet conclusively state that this occurs,
since we have not obtained the continuum limit of the anomaly.

Upon coupling to eleven-dimensional supergravity, a 't Hooft
anomaly in the $SU(2)$ R-symmetry would become an anomaly in
diffeomorphisms of the normal bundle.  This should presumably be
cancelled by a diffeomorphism anomaly due to Chern-Simons terms in
the supergravity action in the presence of magnetic (M5-brane)
sources. We will briefly comment on the contribution of
Chern-Simons terms.

A similar anomaly is known to exist for the $Spin(5)$ R-symmetry
of the six-dimensional $(2,0)$ theory describing parallel
M5-branes \cite{FHMM,HMM,Witteneff,bonorarinaldi,LMT,BHR}. The
anomaly can be directly calculated in the abelian $(2,0)$ theory,
which was first done in \cite{Witteneff}. However, for multiple
M5-branes the anomaly has only been indirectly calculated from the
assumption of anomaly cancellation in M-theory \cite{FHMM,HMM}.
For $N$ M5-branes, the anomaly coefficient is proportional to
$N^3$ at large $N$,  which is consistent with the Weyl anomaly
\cite{henningsonskenderis} and black hole entropy
\cite{HMM,KlebTseytlin,MaldStromWit} calculations. A direct
calculation of the anomaly based on (de)construction should in
principle be possible, but seems to be difficult because the
$SO(5)$ R-symmetry is realized only in the $k\rightarrow\infty$
limit. In the case of intersecting five-branes, the R-symmetry is
realized even for finite $k$, making the anomaly calculation more
tractable.

The organization of this paper is as follows. In section 2 we
review the theory of the D3-D3 intersection in flat space, which
was discussed in \cite{CEGK}. This action is presented in $(2,2)$
superspace. In section 3 we find the quiver diagram for the D3-D3
intersection at a $\CC^2/\ZZ_k$ orbifold, and present the action
in $(2,0)$ superspace.  In section 4 we show how the theory
corresponding to the M5-M5 intersection arises in an appropriate
$k\rightarrow \infty$ limit.   We identify the strings localized
at the M5-M5 intersection. Moreover, we identify the moduli and
R-symmetries of the M5-M5 intersection in the quiver theory, and
discuss the 't Hooft anomaly in the $SU(2)$ R-symmetry. In section
5 we suggest some open problems.

\section{D3-branes intersecting in flat space}
\label{flat} \setcounter{equation}{0}

Before writing the action of intersecting D3-branes at a
$\CC^2/\ZZ_k$ orbifold, it is useful to first write the action of
intersecting D3-branes in flat space.  We shall consider a stack
of $N$ parallel D3-branes in the directions $0123$ intersecting an
orthogonal stack of $N^{\prime}$ D3$^{\prime}$-branes in the
directions $0145$. The action has the form \begin{align} S= S_{\rm D3}
+ S_{\rm D3^{\prime}} + S_{\rm D3-D3^{\prime}}\, .\end{align} The
components $S_{\rm D3}$ and $S_{\rm D3}$ each correspond to a four
dimensional ${\cal N} =4$ theory.  The term $S_{\rm D3-D3^{\prime}}$
contains couplings to a two-dimensional $(4,4)$ hypermultiplet,
leaving only $(4,4)$ supersymmetry unbroken.  The action was
explicitly constructed in $(2,2)$ superspace in \cite{CEGK}, to
which we refer the reader for a more detailed discussion.

It is convenient to define the coordinates
\begin{gather}
z^\pm = X^0 \pm X^1 \, , \qquad x= X^2 + i X^3 \, , \qquad y= X^4
+ i X^5 \, .
\end{gather}
The two-dimensional $(2,2)$ superspace is spanned by $(z,\bar
z,\theta^+, \theta^-, \bar\theta^+, \bar\theta^-)$.  The
four-dimensional fields corresponding to D3-D3 strings are
described by  $(2,2)$ superfields with extra continuous labels
$x,\bar x$, while fields associated to the
D$3^{\prime}$-D$3^{\prime}$ strings have the extra labels $y,\bar
y$. Although the four-dimensional parts of the action will look
strange in $(2,2)$ superspace,  this notation makes sense since
only a two-dimensional supersymmetry is preserved.\footnote{The
procedure of writing supersymmetric $d$-dimensional theories in
terms of a lower dimensional superspace has been discussed in
various places \cite{EGK,CEGK,HG,Hebecker}. }  The fields
associated with D3-D$3^{\prime}$ strings are trapped at the
intersection and have no extra continuous label.

Let us first consider $S_{\rm D3}$, which involves $(2,2)$ superfields
of the form $F(z^+,z^-,\theta,\bar\theta | x,\bar x)$. The
required superfields are a vector superfield $V$, together with
three adjoint chiral superfields $Q_1, Q_2$ and $\Phi$. The gauge
connections $A_{0,1}$ of the $(2,2)$ vector multiplet and the
complex scalar $\phi$ of the $(2,2)$ chiral field $\Phi$ combine
to give the four gauge connections of the four-dimensional ${\cal
N} = 4$ theory.  From $V$ one can build a twisted chiral (field
strength) multiplet,
\begin{align}
  \Sigma \equiv \left\{ \bar \D_+, \D_-\right\}\,,\quad\bar\D_+ = e^{-V}\bar
  D_+ e^{V}\,,\quad \D_- = e^{V} D_- e^{-V}\,,
\end{align}
satisfying $\bar \D_+ \Sigma = \D_- \Sigma=0$.  The scalar
components of $\Sigma, Q_1$ and  $Q_2$ combine to give the six
adjoint scalars of the four-dimensional ${\cal N} =4$ theory. The
field content of the second D3-brane (D$3^{\prime})$ is identical
to that of the first D3-brane with the replacements
\begin{gather}
x \rightarrow y \,,\quad
V \rightarrow {\cal V}\,,\quad
\Sigma \rightarrow \Omega \,,\quad
Q_i \rightarrow S_i \,,\quad
\Phi \rightarrow \Upsilon  \,.
\end{gather}
The fields corresponding to D3-D$3^{\prime}$ strings are the
chiral multiplets $B$ and $\tilde B$ in the $(N, \bar N^{\prime})$
and $(\bar N, N^{\prime})$ representations of the $SU(N) \times
SU(N^{\prime})$ gauge group.  Together they form a $(4,4)$
hypermultiplet.

The components of the action are as follows:
\begin{align}\ \label{bulkaction1}
S_{\rm D3} = \,&\frac{1}{g^2} \int d^2z  d^2x d^4\theta {\rm\, tr}
\left(\Sigma^{\dagger}\Sigma + (\partial_{x} +  g\bar \Phi) e^{gV}
(\partial_{\bar x} + g\Phi) e^{-gV}
+ \sum_{i=1,2} e^{-gV} \bar Q_i e^{gV} Q_i \right) \nonumber\\
+ & \int d^2 z d^2x  d^2 \theta  \epsilon_{ij} {\rm\, tr\,} Q_i
[\partial_{\bar x} + g \Phi, Q_j] + c.c \quad,
\end{align}
\begin{align} \label{bulkaction2}
S_{\rm D3^{\prime}}=\,&\frac{1}{g^2}\int d^2z  d^2y
d^4\theta {\rm\,tr} \left( \Omega^{\dagger}\Omega +
(\partial_{y} + g \bar \Upsilon) e^{g{\cal V}} (\partial_{\bar y} +
g\Upsilon)
e^{-g{\cal V}}
+ \sum_{i=1,2} e^{-g{\cal V}} \bar S_i e^{g{\cal V}} S_i \right) \nonumber\\
+ & \int d^2 z d^2y  d^2 \theta \epsilon_{ij} {\rm
\,tr\,} S_i [\partial_{\bar y} + g\Upsilon, S_j] + c.c \quad,
\end{align}
\begin{align} \label{fullimpaction}
S_{\rm D3- D3^{\prime}} = &\int d^2z d^4 \theta {\rm\, tr}
\left(e^{-g \cal V}\bar B e^{gV} B + e^{-gV} \bar {\tilde B}
e^{g \cal V} \tilde B \right) \nonumber \\
+ & \frac{ig}{{2}} \int d^2z d^2\theta {\,\rm tr}\left(
  B \tilde B Q_1 - \tilde B B S_1\right) + c.c. \quad,
\end{align}
with $d^4\theta= \frac{1}{4} d\theta^+ d\theta^- d\bar\theta^+
d\bar\theta^-$ and $d^2\theta = \frac{1}{2} d \theta^+ d
\theta^-$.

The fact that $S_{\rm D3}$ (or $S_{\rm D3^{\prime}}$) describe
theories with four-dimensional Lorentz invariance, namely ${\cal
N} =4$ super Yang-Mills, can be seen in component notation after
integrating out auxiliary fields. For instance the kinetic term
${\rm tr}\,\partial_\mu \bar q_1
\partial^{\mu} q_1$ with $\mu = 0,1,2,3$, arises from a
combination of the $(2,2)$ K$\ddot{{\rm a}}$hler term ${\rm tr}\,
\bar Q_1 Q_1$ and the superpotential term ${\rm tr}\, Q^1
\partial_z Q^2$.

The fields $B$ and $\tilde B$ acquire masses from expectation
values for $s_1 - q_1$, the lowest component of the superfield $S_1 -
Q_1$, as well as from expectation values for
$\sigma - \sigma^{\prime}$, where $\sigma$ and $\sigma^{\prime}$
are the complex scalars in $V$ (or $\Sigma$) and $\cal V$ (or
$\Omega$). The scalar $s_1 - q_1$   describes fluctuations in the $u =X_6+i
X_7$ direction, and $\sigma-\sigma'$ describes fluctuations in the
$w=X_8+iX_9$ direction. Both $u$ and $w$ are
transverse to both stacks of D3-branes.  On the
Higgs branch the orthogonal D3-branes intersect and the scalars
$s_1-q_1$ and $\sigma-\sigma'$ vanish.
The scalar components $b$ and $\tilde b$
of $B$ and $\tilde B$ have (classical)
expectation values on the Higgs branch.  The scalar components $s_2$
and $q_2$ of $S_2$ and $Q_2$ also have expectation values given by the
vanishing of the F-terms for $S^1$ and $Q^1$:
\begin{align}
\frac{\partial W}{\partial q_1} = \partial_{\bar x} q_2 - g\delta^2(x) b
\tilde b &=0 \,,\qquad
\frac{\partial W}{\partial s_1}= \partial_{\bar y} s_2 -
g\delta^2(y) \tilde b b = 0\,. \label{holomeq}\end{align}
With the geometric identifications $q_2 \sim y/\alpha^{\prime}$ and
$s_2 \sim x/\alpha^{\prime}$, the solutions of these equations
give rise to holomorphic curves\footnote{The holomorphic curves on
the Higgs branch were obtained in discussions with Robert
Helling.} of the form $x y = c\alpha^{\prime}$, when $2\pi i c= g
b\tilde b = g \tilde b b$.

The geometric symmetries of the D3-D3 intersection are as follows.
There is an $SU(2)_L \times SU(2)_R$ R-symmetry corresponding
to rotations in the $6789$ directions transverse to all D3-branes.
Additionally there are $U(1)$ symmetries corresponding to
rotations in the $23$ and $45$ (or $x $ and $y$) planes.  The
charges of the various fields under these symmetries are
summarized in table 1.
\begin{table}[ht]
\begin{center}
\begin{tabular}{cccllcc}
$(4,4)$& $(2,2)$ &  $(2,0)$& components  & $(j_L, j_R)$ & $J_{23}$ &
$J_{45}$ \\
\hline & & & $\sigma, q_1$ &
$(\frac{1}{2},\frac{1}{2})$ &$0$&$0$ \\
Vector& $Q_1, \Sigma$ & $Q_1, \Lambda^{Q_1}_-$ &  $\psi_{q_1}^+,
\bar\lambda_{\sigma}^+$ &
$(0,\frac{1}{2})$ &$\frac{1}{2}$& $-\frac{1}{2}$ \\
& & $\Theta_V, V$&  $\psi_{q_1}^-, \bar\lambda_{\sigma}^-$ &
$(\frac{1}{2},0)$ &$ \frac{1}{2}$&$-\frac{1}{2}$ \\
& & & $v_0, v_1$ & $(0,0)$ &$0$&$0$ \\
\hline
  & & & $\phi$ & $(0,0)$ &$-1$&$0$ \\
Hyper& $Q_2, \Phi$ &  $Q_2, \Lambda_-^{Q_2}$& $ q_2$ & $(0,0)$&$0$&$1$ \\
& & $\Phi, \Lambda_-^{\Phi}$ & $\psi_\phi^+, \bar\psi_{q_2}^+$ & $(\frac{1}{2},0)$ &
$-\frac{1}{2}$ & $-\frac{1}{2}$  \\
& & & $\psi_\phi^-, \bar\psi_{q_2}^-$ & $(0,\frac{1}{2})$ &
$-\frac{1}{2}$&$-\frac{1}{2}$ \\
\hline
& & & $b$ & $(0,0)$ & $-\frac{1}{2}$ & $\frac{1}{2}$  \\
Hyper& $B, \tilde B$ & $B, \Lambda^{B}_-$ & $\tilde b$ & $(0,0)$ &
$-\frac{1}{2}$ &
$\frac{1}{2}$  \\
& & $\tilde B, \tilde \Lambda^{\tilde B}_-$ & $\psi_b^+,
\bar\psi_{\tilde b}^+$ & $(\frac{1}{2},0)$ & $0$ & $0$  \\
& & & $\psi_b^-, \bar\psi_{\tilde b}^-$ & $(0,\frac{1}{2})$ & $0$
& $0$  \\
\hline & & &$\omega, s_1$ &
$(\frac{1}{2},\frac{1}{2})$ & $0$ &$0$ \\
Vector& $S_1, \Omega$ & $S_1, \Lambda^{S_1}_-$ & $\psi_{s_1}^+,
\bar\psi_{\omega}^+$ &
$(0,\frac{1}{2})$ & $\frac{1}{2}$ & $-\frac{1}{2}$  \\
& & $\Theta_{\cal V}, {\cal V}$ & $\psi_{s_1}^-,
\bar\psi_{\omega}^-$ &
$(\frac{1}{2},0)$ & $\frac{1}{2}$ & $-\frac{1}{2}$ \\
& & & $a_0, a_1$ & $(0,0)$ & $0$ &$0$  \\
\hline
& & & $\upsilon$ & $(0,0)$ & $0$ & $1$  \\
Hyper& $S_2, \Upsilon$ & $S_2, \Lambda_-^{S_2}$ & $ s_2$ & $(0,0)$ &
$-1$ & $0$  \\
& & $\Upsilon, \Lambda_-^\Upsilon$ & $\lambda_\upsilon^+,
\bar\psi_{s_2}^+$ & $(\frac{1}{2},0)$ & $\frac{1}{2}$ &
$\frac{1}{2}$ \\
& & & $\lambda_\upsilon^-, \bar\psi_{s_2}^-$ & $(0,\frac{1}{2})$ &
$\frac{1}{2}$ & $\frac{1}{2}$ \\
\end{tabular}
\caption{Field content of the D3-D3 intersection.}\label{rsym}
\end{center}
\end{table}
The $U(1)$ symmetries generated by $J_{45}$ and $J_{23}$ are
manifest in $(2,2)$ superspace.  The $U(1)$ generated by $J_{45}$
has the following action:
\begin{align}\label{j45}
&\theta^{+} \rightarrow e^{i\alpha/2} \theta^{+}\,, &&B\rightarrow
e^{i \alpha/2} B\,, &&
Q_2 \rightarrow e^{i\alpha} Q_2 \,,&&&&\nonumber\\
&\theta^{-} \rightarrow e^{i\alpha/2} \theta^{-}\,, &&\tilde B
\rightarrow e^{i \alpha/2} \tilde B\,,  &&
\Upsilon \rightarrow  e^{+i\alpha} \Upsilon \,, \nonumber\\
&y\rightarrow e^{i\alpha} y\,,
\end{align}
with all remaining fields being singlets. The $U(1)$ generated by
$J_{23}$ acts as
\begin{align}\label{j23}
&\theta^{+} \rightarrow e^{-i\alpha/2} \theta^{+}\,,
&&B\rightarrow e^{-i \alpha/2} B\,, &&
S_2 \rightarrow e^{-i\alpha} S_2\,,&&&& \nonumber \\
&\theta^{-} \rightarrow e^{-i\alpha/2} \theta^{-}\,, &&\tilde B
\rightarrow e^{-i \alpha/2} \tilde B\,,  &&
\Phi\rightarrow e^{-i\alpha}\Phi\,, \nonumber \\
&x\rightarrow  e^{-i\alpha} x\,.
\end{align}

\section{D3-D3 intersection at a $\CC^2/\ZZ_k$ orbifold}
\setcounter{equation}{0}

To (de)construct the theory of the M5-M5 intersection,  we shall
consider a pair of intersecting stacks of D3-branes at an
$\CC^2/\ZZ_k$ orbifold point. One set of D3-branes is located at
$X^{4,5,6,7,8,9} = 0$ while the other set of ${\rm D3}^{\prime}$
branes is located at $X^{2,3,6,7,8,9}=0$. The $\CC^2/\ZZ_k$ is
spanned by the coordinates $u = X^6 + iX^7$ and $w= X^8 + iX^9$
subject to the orbifold condition $u \sim \xi u, w \sim \xi^{-1}
w$ where $\xi = \exp(2\pi i/k)$.  Before orbifolding,  the theory
of intersecting D3-branes has $(4,4)$ supersymmetry with an
$SU(2)_L \times SU(2)_R$ R-symmetry. The $SU(2)_L \times SU(2)_R$
component of the R-symmetry acts as an $SO(4)$ transformation on
the real components of $u$ and $w$, which are the coordinates
$X^{6,7,8,9}$. The orbifold breaks $SU(2)_L \times SU(2)_R$  to
$SU(2)_L$, under which the pair $u,w^*$ transform as a doublet.
Moreover, the supersymmetry is broken from  $(4,4)$ to $(4,0)$.
The chiral nature of this theory will prove important later when
we find evidence that tensionless strings give rise to a 't Hooft
anomaly in an R-symmetry of the M5-M5 intersection.

\subsection{Orbifold projection}
\begin{figure}[!ht]
\begin{center}
\includegraphics[height=5cm,clip=true,keepaspectratio=true]{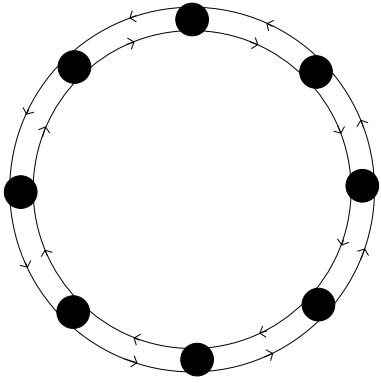}
\caption{Quiver diagram for parallel D3-branes at a $\CC^2/\ZZ_k$
orbifold (here $k=8$). Each node corresponds to an ${\cal N} =2$ vector
multiplet, while double lines between neighboring nodes correspond
to an $\N=2$ hypermultiplet.}\label{quiv}
\end{center}
\end{figure}

The Lagrangian describing the D3-D3 intersection in the
$\CC^2/\ZZ_k$ orbifold can be obtained from the action of the
D3-D3 intersection in flat space given in
(\ref{bulkaction1}) - (\ref{fullimpaction}). Following
refs.~\cite{DM,JM} we start with $Nk$
D3-branes intersecting $N^\prime k$ D3-branes in a flat
background and project out the degrees of freedom which are not
invariant under the $\ZZ_k$ orbifold group, which is generated by
a combination of a gauge symmetry and an R-symmetry. An important
constraint on the orbifold action is that the theory on each stack
of D3 branes (ignoring strings connected to the other stack)
should be the ${\cal N} =2, d=4$ super Yang-Mills theory described
by the quiver in figure \ref{quiv}, with gauge group $SU(N)^k$ or
$SU(N^{\prime})^k$.

The orbifold action which gives the quiver of figure 1 for both
the D3 and the D3$^{\prime}$ degrees of freedom separately, and
breaks the $SU(2)_L \times SU(2)_R$ R-symmetry to $SU(2)_L$ is as
follows. The embedding of the $\ZZ_k$ orbifold group in the
$U(Nk)$ and $U(N^{\prime}k)$ gauge groups is given by
\begin{align}
g(\xi) &= \begin{pmatrix} I_{N\times N} & & &\\ & \xi I_{N\times
N} & &
\\ & & \xi^2I_{N\times N} & \\
& & & \ddots \label{g}
\end{pmatrix} \,,\\ g^{\prime}(\xi) &=
\begin{pmatrix} I_{N^{\prime}\times N^{\prime}} & & &\\
& \xi I_{N^{\prime}\times N^{\prime}} & &
\\ & & \xi^2I_{N^{\prime}\times N^{\prime}} & \\
& & & \ddots\label{g'}
\end{pmatrix} \,,
\end{align}
where $\xi$ is the generator $\exp(2\pi i/k)$ of $\ZZ_k$. The
embedding of the $\ZZ_k$ orbifold group in the R-symmetry is given
by
\begin{align}
h(\xi) = e^{i\pi \sigma^3/k} \,,
\end{align}
where $h$ belongs to $SU(2)_R$. The field theory describing the
D3-D$3^{\prime}$ intersection at the orbifold is then obtained
from that of the D3-D$3^{\prime}$ intersection in flat space by
projecting out fields which are not invariant under the orbifold
action. The result is an $SU(N)^k \times SU(N^{\prime})^k$ gauge
theory with $(4,0)$ supersymmetry and $SU(2)_L \times U(1)$
R-symmetry.

In $(2,2)$ superspace, the orbifold acts on superspace coordinates as
\begin{align}\label{supe}
&\theta^- \rightarrow \xi\theta^- \,,
\end{align}
but trivially on $\theta^+$. On the $(2,2)$ superfields the orbifold
acts as
\begin{align}
&{\rm  D3}:  &&V\rightarrow g(\xi)Vg^{\dagger}(\xi)\,,
&&\Sigma \rightarrow \xi^{-1}g(\xi)\Sigma g^{\dagger}(\xi)\,,&&\nonumber \\
&&&Q_1 \rightarrow \xi g(\xi)Q_1g^{\dagger}(\xi)\,,  &&Q_2
\rightarrow g(\xi)Q_2g^{\dagger}(\xi) \,,&&\nonumber \\
&&&\Phi \rightarrow g(\xi)\Phi g^{\dagger}(\xi)\,,&&\nonumber\\
&{\rm D3}^{\prime}:&&{\cal V}\rightarrow
g^{\prime}(\xi){\cal V}g^{\prime\dagger}(\xi)\,,  &&\Omega
\rightarrow \xi^{-1}g^{\prime}(\xi)\Omega
g^{\prime\dagger}(\xi)\,,&&\nonumber\\
&&&S_1 \rightarrow \xi g^{\prime}(\xi)S_1 g^{\prime\dagger}(\xi)\,,
&&S_2 \rightarrow g^{\prime}(\xi)S_2g^{\prime\dagger}(\xi)\,,
&&\nonumber \\
&&&\Upsilon \rightarrow g^{\prime}(\xi)\Upsilon
g^{\prime\dagger}(\xi)\,,
&&\nonumber\\
&{\rm D3-D3^{\prime}}:&&B\rightarrow
g(\xi)Bg^{\prime\dagger}(\xi)\,, && \tilde B \rightarrow
g^{\prime}(\xi) \tilde B g^{\dagger}(\xi)\,, \label{orbac}
\end{align}
with $g$, $g'$ as in (\ref{g}), (\ref{g'}).
Starting with the action (\ref{bulkaction1}) -
(\ref{fullimpaction}) and projecting out the degrees of freedom
which are not invariant under (\ref{orbac})  will give a $(4,0)$
supersymmetric action with manifest $(2,0)$ supersymmetry.

To illustrate how the orbifold acts on components, we consider the
action (\ref{orbac}) on the $(2,2)$ twisted superfield $\Sigma$. On the
bosonic components, this corresponds to
\begin{align}
\sigma &\sim \xi^{-1} g(\xi) \sigma g^{\dagger}(\xi)\,,\quad
F_{01} \sim g(\xi) F_{01} g^{\dagger}(\xi) \,.
\end{align}
This is consistent with the fact that the field $\sigma$
characterizes fluctuations transverse to both D3-branes, i.e.\
fluctuations in the orbifold directions. This field is naturally
associated with fluctuations in the $w= X^8+ i X^9$ directions
which satisfy the orbifold condition $w \sim \xi^{-1} w$.   Upon
projecting out the parts which are not invariant under the
orbifold, $\sigma$ becomes a set of $k$ bifundamentals in the
representations $( \cdots N, \bar N, \cdots)$ of $SU(N)^k$.  These
bifundamental fields are written as $\sigma_{j,j+1}$,  where $j=1
\cdots k$ and the first(second) index labels the gauge group with
respect to which the field is a fundamental (antifundamental).
Fields which are adjoints with respect to one of the factors will
be written with a single index.

\subsection{Quiver action in two-dimensional $(2,0)$ superspace}

Since the $(4,4)$ supersymmetry of the action (\ref{bulkaction1}) -
(\ref{fullimpaction}) is broken down to $(4,0)$ by the orbifold, an adequate
formulation of the corresponding quiver gauge theory is best given in
$(2,0)$ superspace. In order to project out the degrees of freedom which are
not invariant under the orbifold, we rewrite the parent action using
manifest
$(2,0)$ supersymmetry. To this end, we decompose the $(2,2)$ superfields
under
$(2,0)$ supersymmetry.  The decomposition of the $(2,2)$
superfields is as follows (see for instance \cite{Witten,Garcia}):\\

i) $(2,2)$ vector $\rightarrow$ $(2,0)$ vector $+$
$(2,0)$
chiral,

ii) $(2,2)$ chiral $\rightarrow$ $(2,0)$ chiral $+$
$(2,0)$ Fermi.\\

\noindent These $(2,0)$ superfields have the following component
decomposition:  \\

\begin{minipage}{14.5cm}
\noindent i) $(2,0)$ vector superfield $V$:  two gauge connections
$A_0, A_1$ and one fermion $\chi_-$,

\noindent ii) $(2,0)$ chiral superfields  $\Phi$: one complex
scalar $\phi$ and a fermion $\psi_+$,

\noindent iii) $(2,0)$ fermi superfields $\Lambda$: one chiral
fermion $\lambda_-$.  The full expansion of this anticommuting superfield
contains an auxiliary field and a holomorphic function of $(2,0)$ chiral
superfields. \\
\end{minipage}

\begin{figure}[!ht]
\begin{center}
\includegraphics[height=18cm,clip=true,keepaspectratio=true]{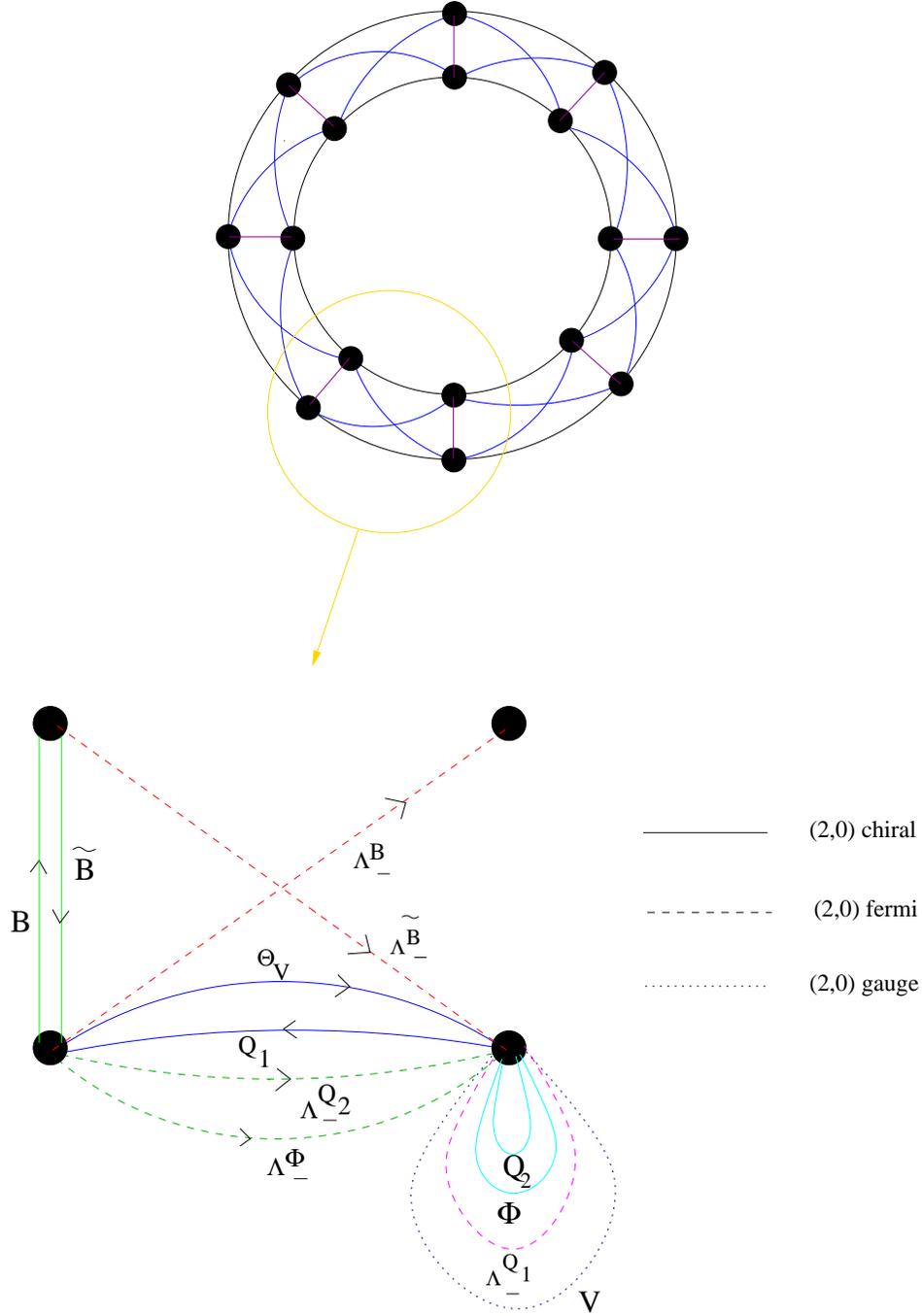}
\caption{``Mating Moose'': Quiver diagram for intersecting D3-branes at a
$\CC^2/\ZZ_k$ orbifold (with $k$=8).  The nodes of the inner and
outer circle are associated with the $SU(N')^k$ and $SU(N)^k$
gauge groups respectively.  The parts which have not been drawn in
the detailed ``close-up'' are easily inferred from the $\ZZ_k$
symmetry and by swapping D3 degrees of freedom with D$3^{\prime}$
degrees of freedom.}\label{superquiver}
\end{center}
\end{figure}

\noindent For the theory given by the action (\ref{bulkaction1}) -
(\ref{fullimpaction}), the decomposition of the $(2,2)$ superfields of the
D3-D3 intersection in flat space gives the following $(2,0)$ superfields (we
shall henceforward write $(2,2)$ superfields in boldface):
\begin{align}
&{\rm D3}:  &&{\bf Q}_1 \rightarrow Q_1, \Lambda^{Q_1}, && {\bf
Q}_2 \rightarrow
  Q_2,\Lambda^{Q_2}, && {\bf\Phi} \rightarrow \Phi, \Lambda^\Phi, &&
  {\bf V} \rightarrow V, \Theta_V\,,&&\nonumber
  \\
&{\rm D3}^{\prime}: &&{\bf S}_1 \rightarrow S_1, \Lambda^{S_1}, &&
{\bf S}_2 \rightarrow S_2,
  \Lambda^{S_2} \,,&& {\bf\Upsilon} \rightarrow \Upsilon,
  \Lambda^\Upsilon, &&
  {\bf {\cal V}} \rightarrow {\cal V}, \Theta_{\cal V}\,,&&\nonumber\\
&{\rm D3}-{\rm D3}: &&{\bf B} \rightarrow B, \Lambda^B, && \tilde {\bf
B} \rightarrow \tilde
  B, \Lambda^{\tilde B}\,.
\end{align}
Since we wish to obtain the action for the D3-D3 intersection at
the $\CC^2/\ZZ_k$ singularity in $(2,0)$ superspace, we
write the orbifold action (\ref{supe}), (\ref{orbac}) in $(2,0)$
superspace. In terms of the $(2,0)$ decomposition, the orbifold
acts as follows:
\begin{align} \label{orbifoldconstraints}
&{\rm D3}:&& Q_1 \rightarrow \xi g(\xi)
Q_1g^{\dagger}(\xi)\,,\quad &&\Lambda^{Q_1} \rightarrow g(\xi)
\Lambda^{Q_1} g^{\dagger}(\xi)\,, &&\nonumber
\\
&&&Q_2 \rightarrow g(\xi) Q_2 g^{\dagger}(\xi)\,,\quad &&\Lambda^{Q_2}
\rightarrow \xi^{-1}g(\xi)
\Lambda^{Q_2} g^{\dagger}(\xi)\,, &&\nonumber\\
&&&\Phi \rightarrow g(\xi)\Phi g^{\dagger}(\xi)\,,\quad
&&\Lambda^{\Phi}
\rightarrow \xi^{-1} g(\xi) \Lambda^{\Phi} g^{\dagger}(\xi)\,, &&\nonumber\\
&&&V \rightarrow g(\xi) V g^{\dagger}(\xi)\,,\quad &&\Theta_V \rightarrow
\xi^{-1} g(\xi)\Theta_V g^{\dagger}(\xi)\,,&&\nonumber\\
\nonumber\\
&{\rm D3}^{\prime}:&& S_1 \rightarrow \xi g'(\xi)
S_1g'^{\dagger}(\xi)\,,\quad &&\Lambda^{S_1} \rightarrow g'(\xi)
\Lambda^{S_1} g'^{\dagger}(\xi)\,,
&&\nonumber\\
&&&S_2 \rightarrow g'(\xi) S_2 g'^{\dagger}(\xi)\,,\quad
&&\Lambda^{S_2} \rightarrow \xi^{-1}g'(\xi) \Lambda^{S_2}
g'^{\dagger}(\xi)\,, &&\nonumber\\
&&&\Upsilon \rightarrow g'(\xi)\Upsilon g'^{\dagger}(\xi)\,,\quad
&&\Lambda^{\Upsilon}
\rightarrow \xi^{-1} g'(\xi) \Lambda^{\Upsilon} g'^{\dagger}(\xi)\,,
&&\nonumber\\
&&&{\cal V} \rightarrow g'(\xi) {\cal V} g'^{\dagger}(\xi)\,,\quad
&&\Theta_{\cal V}\rightarrow \xi^{-1} g'(\xi)\Theta_{\cal V}
g'^{\dagger}(\xi)\,,&&\nonumber\\
\nonumber\\
&{\rm D3}-{\rm D3}^{\prime}:&& B \rightarrow g(\xi) B
g'{}^\dagger (\xi) \,,\quad &&\Lambda^{B} \rightarrow \xi^{-1}
g(\xi) \Lambda^{B}
g'{}^\dagger(\xi)  \,,\quad  &&\nonumber\\
&&&\tilde B \rightarrow g'(\xi) \tilde B g{}^\dagger (\xi) \,,\quad
&&\Lambda^{\tilde B} \rightarrow \xi^{-1} g'(\xi) \Lambda^{\tilde B}
g^\dagger(\xi)  \,.&&
\end{align}
Each component of a $(2,0)$ superfield transforms under the orbifold action
in
the same way as the $(2,0)$ superfield itself. Note that this was not the
case
for $(2,2)$ superfields.  The degrees of freedom which are invariant under
(\ref{orbifoldconstraints}) together with their $SU(N)^k \times
SU(N^{\prime})^k$ gauge transformation properties are summarized by the
quiver
diagram in figure~\ref{superquiver}. The quiver consists of an inner and an
outer ring. Each of them is equivalent to the moose shown in figure
\ref{quiv}
which provides the field content for the (de)construction of the
six-dimensional $(2,0)$ superconformal field theory. We will see below that
the spokes in the diagram, which connect both rings, represent the degrees
of
freedom for the (de)construction of a $\N=2, d=4$ field theory located at
the
M5-M5 intersection.

\newpage

We do not need the full action of the $(4,0)$ quiver theory. For now we just
give the the $(2,0)$ term analogous to a superpotential, which will be all
that we require for most purposes.  Superpotentials of $(2,0)$ theories have
the generic structure
\begin{align}
W= \int d\theta^+ \sum_a \Lambda^a J_a(\Phi_i)|_{\theta^+ = 0} \,,
\end{align}
where $J_a(\Phi_i)$ is a holomorphic function of the chiral
superfields satisfying a certain constraint (see the appendix).
For the D3-D3 intersection at a $\CC^2/\ZZ_k$ orbifold, this term
descends from the superpotential of the D3-D3 intersection in
flat space which is presented in $(2,0)$ superspace
in the appendix. Upon projecting out the degrees of freedom which are
not invariant under the orbifold (\ref{orbifoldconstraints}), one
obtains the $(2,0)$ superpotential
\begin{align}
W=W_{\rm D3} + W_{\rm D3'} + W_{\rm D3-D3'}\,, \label{super}
\end{align}
where
\begin{align}
W_{\rm D3}=  \int d^2x d\theta^+ {\rm tr}_{N\times N}& \left(g
\Lambda^\Phi_{j,j+1}  ( Q^2_{j+1} Q^1_{j+1,j} \,- \,
   Q^1_{j+1,j} Q^2_{j})\right. \label{super1}\\
+ &\,\Lambda^{Q^1}_j[\partial_{\bar x}+g\Phi_{j},Q^2_{j}]\nonumber\\
+ &\, \left.\left.\Lambda^{Q^2}_{j,j+1}( - \partial_{\bar
x}Q^1_{j+1,j} - g Q^1_{j+1,j} \Phi_j
+ g \Phi_{j+1} Q^1_{j+1,j}) \right)\right\vert_{\bar\theta^+ = 0}
\,,\nonumber\\
W_{\rm D3'}= \int d^2y
d\theta^+ {\rm tr}_{N^{\prime}\times N^{\prime}}& \left(g
\Lambda^\Upsilon_{j,j+1}  ( S^2_{j+1} S^1_{j+1,j} \,- \,
   S^1_{j+1,j} S^2_{j})\right. \label{super2} \\
+ &\,\Lambda^{S^1}_j[\partial_{\bar x}+g\Upsilon_{j},S^2_{j}]\nonumber\\
+ &\,\left.\left. \Lambda^{S^2}_{j,j+1}( - \partial_{\bar
x}S^1_{j+1,j} - g S^1_{j+1,j} \Upsilon_j
+ g \Upsilon_{j+1} S^1_{j+1,j}) \right)\right\vert_{\bar\theta^+ = 0}
\,,\nonumber\\
W_{\rm D3-D3'} = g \int d\theta^+ {\rm tr}_{N\times N}& \left(
\Lambda^B_{j,j+1}(\tilde B_{j+1}Q^1_{j,j+1} - S^1_{j+1,j}
\tilde B_j) + \Lambda^{Q^1}_j B_j\tilde B_j \right) \label{superpotential}
\\
+ {\rm tr}_{N^{\prime} \times N^{\prime}} & \left.\left(
\Lambda^{\tilde B}_{j,j+1} (Q^1_{j+1,j}B_j - B_{j+1} S^1_{j+1,j})
- \Lambda^{S^1}_j \tilde B_j B_j \right)\right\vert_{\bar\theta^+
= 0} \,.\nonumber
\end{align}

In order to see that this theory has indeed $(4,0)$ supersymmetry, we record
the basic structure of the $(4,0)$ multiplets which appear. These are as
follows:\\
\\
\noindent i) $(4,0)$ hypermultiplets composed of two $(2,0)$
chiral multiplets: There are five multiplets of this type
containing the pairs $(B,\tilde B), (\Phi,Q_2), (\Theta_V, Q_1),
(\Upsilon, S_2)$
and $(\Theta_{\cal V}, S_1)$. \\

\noindent ii) $(4,0)$ vector multiplets composed of one $(2,0)$
vector multiplet and one $(2,0)$ fermi multiplet: There are two
multiplets of this type containing the pairs $(V,
\Lambda^{Q_1})$ and $({\cal V}, \Lambda^{S_1})$.\\

\noindent iii) $(4,0)$ Fermi multiplets composed of
one\footnote{There is no need to add degrees of freedom to make a
$(4,0)$ Fermi multiplets out of a $(2,0)$ Fermi multiplet \cite{Garcia}.}
$(2,0)$ Fermi multiplet:  There are six multiplets of this type
corresponding to the $(2,0)$ fermi multiplets $\Lambda^B,
\Lambda^{\tilde B}, \Lambda^{\Phi}, \Lambda^{Q_2},
\Lambda^{\Upsilon}$ and
$\Lambda^{S_2}$. \\
\\
\noindent The transformation properties under the $(4,0)$
$SU(2)_L$ R-symmetry are readily obtained from table 1 on page
\pageref{rsym}.  Note that the $SU(2)_L$ R-symmetry acts on the
degrees of freedom of either the inner or outer ring of the quiver
diagram as the $SU(2)$ R-symmetry of the associated ${\cal N} =2,
d=4$ theory.

In the following section, we shall make use of this superpotential to
discuss the (de)con\-struction of the M5-M5 intersection.

\section{(De)constructing the M5-M5 intersection}

\setcounter{equation}{0}

\begin{figure}[!h]
\begin{center}
\includegraphics[keepaspectratio, scale=1]{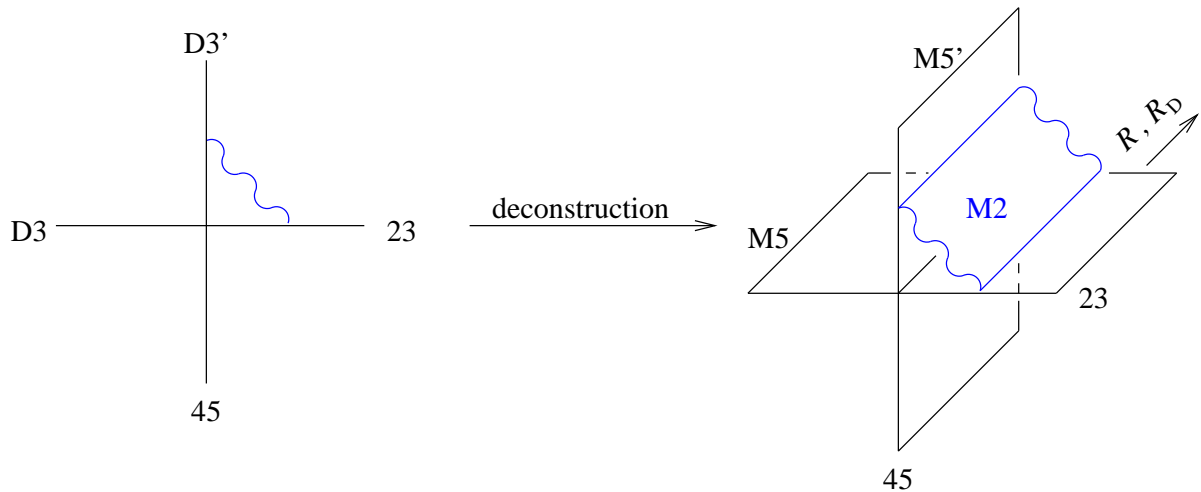}
\caption{(De)construction of two extra dimensions along the torus
with radii $R$ and $R_D$. The common directions $x^0$ and $x^1$ as
well as the four orbifold directions are surpressed.}\label{deconstruction}
\end{center}
\end{figure}

The inner and outer circle of the quiver diagram in figure
\ref{superquiver} are each separately equivalent to the quiver
diagram of figure \ref{quiv}, which (de)constructs the $(2,0)$
theory upon taking the appropriate large $k$ limit
\cite{Arkani-Hamed}. The new twist here is that there are degrees
of freedom connecting the inner and outer rings. These are
localized at the intersection of the D3-branes, and it is natural
to expect that in the large $k$ limit, these correspond to the
tensionless strings localized at the intersection of M5-branes.
One reason to expect this follows from a trivial extension of an
argument given in \cite{Arkani-Hamed} based on the IIB string
theory embedding. The basic idea is that in the $k
\rightarrow\infty$ limit, the $\CC^2/\ZZ_k$ orbifold appears as a flat
$S^1 \times \RR^3$ geometry to D3-brane sufficiently far away from
the orbifold point (or sufficiently far out on the Higgs branch).
For intersecting D3-branes, T-dualizing and lifting to M-theory on
this space gives rise to intersecting M5-branes wrapping a torus
of fixed dimensions. The strings stretched between the orthogonal
D3-branes then become membranes stretched between M5-branes, as
shown in figure \ref{deconstruction}. In
the following, we shall focus on the field theoretic origins of
the tensionless strings at the intersection.

\subsection{(De)constructing the $(2,0)$ theory}

Before discussing the strings localized at the intersection, we
shall briefly review the field theoretic arguments behind the
(de)construction of the six-dimensional $(2,0)$ theory discovered
in \cite{Arkani-Hamed}. The quiver diagram of the deconstructed
theory is that of figure \ref{quiv}, which describes a
superconformal ${\cal N} =2, d=4$ gauge theory with gauge group
$SU(N)^k$. The hypermultiplets described by the double lines
stretched between adjacent nodes contain two complex scalars in
bifundamental representations.  The quiver diagram can be viewed
as a discretization of an extra circular spatial dimension if one
takes all the bifundamental scalars to have the same non-zero
expectation value.  At this point on the Higgs branch the gauge
symmetry is broken from $SU(N)^k$ to the diagonal $SU(N)$.

To make closer contact with our work,  we show how the extra
dimensions arise from the
${\cal N} =2, d=4$ theory using the language of two-dimensional
$(2,0)$ superspace.  Consider the term $W_{\rm D3}$ in the
superpotential (\ref{super}),  which involves only fields on the
outer ring of the quiver diagram. Deconstructing the
six-dimensional $(2,0)$ theory involves going to a particular
point in the Higgs branch of the ${\cal N} =2, d=4$ theory
described by the outer ring.  At this point $\langle q^1_{j+1,j}
\rangle = vI$ for all $j$, where $v$ is real and $q^1$ is the
scalar component of $Q^1$. One then has an effective
superpotential with the quadratic terms
\begin{align}
W_{\rm D3} = \int d^2x d\theta^+ {\rm tr}_{N\times N}& \left(g v
   \Lambda^\Phi_{j,j+1}(Q^2_{j+1}-Q^2_{j})
+  gv\Lambda^{Q^2}_{j,j+1}( \Phi_{j+1} - \Phi_{j} ) \right.
\nonumber \\
&\,+\left.\left.\Lambda^{Q^1}_j\partial_{\bar x}Q^2_{j} -
\Lambda^{Q^2}_{j,j+1}\partial_{\bar x}{\cal P}_{j+1,j}
\right)\right\vert_{\theta^+ = 0} \,, \label{three}
\end{align}
where\footnote{The field ${\cal P}_{j+1,j}$ can be interpreted as
part of a gauge connection in an extra spatial latticized
direction. Terms other than the superpotential must also be
included to see this.} ${\cal P}_{j+1,j} = v- Q^1_{j+1,j}$. The
first and second terms in (\ref{three}) can be viewed as kinetic
terms on a lattice with $k$ sites and lattice spacing $a =
1/{gv}$.  The bosonic kinetic terms arise upon integrating out
auxiliary fields.   From a two-dimensional point of view, the
first two terms in (\ref{three}) give rise to a mass
matrix\footnote{Strictly speaking, we must also include the
contribution to the mass matrix coming from terms other than the
superpotential.  These terms are related to those of the
superpotential by $(4,0)$ supersymmetry, and modify an overall
factor in the mass matrix.  } with eigenvalues
\begin{align}
M^2 = (gv)^2 |e^{2\pi i n/k} - 1|^2 \,. \label{spectrum}
\end{align}
For sufficiently large $k$,  this becomes a Kaluza-Klein spectrum
$M^2 = (n/R)^2$ with $R= \frac{k}{gv}$.

Yet another compact dimension is generated due to the S-duality of
the ${\cal N}=2, d=4$ gauge theory.  Under S-duality $g\rightarrow
k/g$ and one therefore expects a spectrum of S-dual states with
masses
\begin{align}
M_D^2 = \left (\frac{kv}{g} \right)^2|e^{2\pi i n/k} - 1|^2\,.
\end{align}
For large $k$ and fixed $n$ there is a Kaluza-Klein spectrum on an
S-dual circle of radius
\begin{align}
2\pi R_D = \frac{g}{v}\,.
\end{align}
The continuum limit is obtained by taking $k\rightarrow \infty$
with $R, R_D$ fixed.  This requires that one goes to strong
coupling $g\sim \sqrt{k}$ and that one goes far out onto the Higgs
branch $v\sim \sqrt{k}$.

\subsubsection{A note on stability of the spectrum}

The existence of the continuum limit is actually more subtle than
the previous discussion suggests since it includes a strong
coupling limit.  Although string theory indicates the limit should
exist,  a field theoretical argument would have to demonstrate the
validity of the semiclassical spectrum (\ref{spectrum}) at strong
coupling, and $k\rightarrow\infty$ at fixed $n$.  Strictly speaking, 
this spectrum is not a BPS mass formula for finite $k$,  since the
``charge'' $n$ is defined modulo $k$ and is therefore not a
central charge.  Assuming the existence of a continuum limit with 
enhanced supersymmetry, 
the spectrum is BPS with respect to this enhanced 
supersymmetry.   In \cite{CET},  an argument that the supersymmetry enhancement
is robust at low energies was given by studying the Seiberg-Witten
curve of the ${\cal N}=2, d=4$ quiver gauge theory.   

A further argument in favor of the stability of the spectrum at finite $k$ is
as follows.  Although the first two terms in (\ref{three}) are lattice kinetic
terms, they appear in the $(2,0)$ superpotential, which has a
holomorphic structure and is protected against radiative
corrections. If we were to work with four-dimensional ${\cal N}
=1$ superspace,  we would also find that the lattice kinetic terms
arise in part from the effective superpotential on the Higgs
branch.  In ${\cal N}=1, d=4$ superspace,  the superpotential is
\begin{align}
W = g \sum_{j=1}^k {\rm tr}\left( \Phi^1_j \,\Phi^2_{j,j+1}
\Phi^3_{j+1,j} - \Phi^1_{j+1}\Phi^3_{j+1,j}\Phi^2_{j,j+1}\right) \,.
\end{align}
The effective superpotential corresponding to lattice kinetic
terms is obtained on the Higgs branch by setting $\Phi^2_{j,j+1} =
v + \Gamma^2_{j,j+1}$ and $\Phi^3_{j+1,j} = v + \Gamma^3_{j+1,j}$.
The non-renormalization of the effective superpotential and terms
related to it by supersymmetry is crucial for the stability of the
spectrum (\ref{spectrum}) at large $g$, and to the existence of a
continuum limit.

Note that the non-renormalization of the lattice kinetic terms is
somewhat akin to the non-renormalization of the metric on the
Higgs branch of four-dimensional ${\cal N} =2$ gauge theories.  
The latter non-renormalization can be argued, albeit in an unconventional
way, by writing the action in two-dimensional $(2,2)$
superspace. The kinetic terms for the ${\cal N}=2, d=4$
hypermultiplet then arise partially from a $(2,2)$ superpotential
of the form $\epsilon_{ij}Q_i \partial_x Q_j$ as in
(\ref{bulkaction1}).

\subsection{Strings at the intersection}

Let us now consider the same $k\rightarrow \infty$ limit as above
for the case in which there are orthogonal intersecting stacks of
D3-branes.  We will initially take the Higgs branch moduli for the
${\cal N}=2, d=4$ theories on the inner and outer ring of the
quiver to be equal,  such that  $\langle s^1_{j+1,j} \rangle
= vI_{N\times N}$ and $\langle q^1_{j+1,j}\rangle = vI_{N^{\prime}
\times N^{\prime}}$. In this case,
the inner and outer rings of the quiver can be expected to
separately (de)construct the six-dimensional $(2,0)$ theory
compactified on tori with the same dimensions.  However one must
also consider the strings stretching between the D3-branes, i.e.\
the ``spokes'' which connect the inner and outer rings of the
quiver. We shall now argue that these (de)construct tensionless
strings living at a four-dimensional intersection of the two
six-dimensional world volumes.

The ``spoke'' degrees of freedom correspond to the $(2,0)$ chiral
fields $B_j,\tilde B_j$ and Fermi fields ${\Lambda^B}_{j,j+1},
{\Lambda^{\tilde B}}_{j,j+1}$,  which describe strings stretched
between the two stacks of D3-branes. For $\langle s^1_{j+1,j} \rangle=
vI_{N\times N}$ and $\langle q^1_{j+1,j}\rangle = vI_{N^{\prime} \times
N^{\prime}}$,  the quadratic part of the effective superpotential
is
\begin{align}\label{latkin}
W_{\rm D3-D3'} = gv\int d\theta^+ {\rm tr} \left. \left[
\Lambda^{\tilde B}_{j,j+1}(B_j - B_{j+1}) + (\tilde B_{j+1} -
\tilde B_j)\Lambda^B_{j,j+1}\right]\right\vert_{\bar\theta^+ = 0} \,
\end{align}
which follows from (\ref{superpotential}).  This can also be
viewed as a lattice kinetic term.  The same mass matrix arises for
the fundamental degrees of freedom at the intersection as for
those on the inner and outer circles of the quiver. Therefore
these degrees of freedom also carry momentum in an extra dimension
of radius $R$. The full theory is again expected to exhibit
S-duality, based on its embedding in string theory. Thus there
should also be S-dual degrees of freedom at the intersection which
carry momentum in an extra dimension of radius $R_D$. Dyonic
states carry momenta in both extra directions.  The precise nature
of degrees of freedom which are S-dual to the fundamental degrees
of freedom $B, \Lambda^B, \tilde B$ and~$\Lambda^{\tilde B}$
remains an open question at the moment. However, assuming
S-duality, the $k\rightarrow \infty$ limit generates two
six-dimensional world volumes intersecting over four dimensions
from a theory with two four-dimensional world volumes intersecting
over two dimensions. Note that the inner and outer rings of the
quiver do not see independent extra directions, since the apparent
$\ZZ_k \times \ZZ_k$ symmetry is broken to $\ZZ_k$ by couplings to
the degrees of freedom at the intersection.

The spoke degrees of freedom should be interpreted as tensionless
strings wrapping the compact directions rather than particles. To
see this, it is helpful to move the orthogonal stacks of D3-branes
to different points in the orbifold.  This corresponds to going to
different points on the Higgs branches of theories described by
the inner and outer rings of the quiver. For the inner ring the
Higgs branch is characterized by vevs for $\sigma_{j,j+1}$ and
$\bar q^1_{j,j+1}$ which form a doublet $Y_{j,j+1}$ of the
$SU(2)_L$ R-symmetry. Similarly the Higgs branch for the outer
ring is characterized by vevs for $\omega_{j,j+1}$ and
$s^1_{j,j+1}$ which also form a doublet $Y^{\prime}_{j,j+1}$ of
$SU(2)_L$.   Consider the following point in the moduli space:
\begin{align}
Y_{j,j+1} = \begin{pmatrix} v+\Delta/2 \cr v + \Delta/2 \end{pmatrix}
\,,\qquad
Y_{j,j+1}^{\prime} = \begin{pmatrix} v-\Delta/2 \cr v - \Delta/2
\label{sep}
\end{pmatrix}\quad.
\end{align}
where $\Delta$ is real.   One might worry that the extra
dimensions seen by the degrees of freedom on the inner and outer
rings of the quiver are no longer the same, since  at different
points on the Higgs branches, $\langle Y \rangle \ne \langle
Y^{\prime} \rangle$, the radii are apparently different. However
we shall keep $v\Delta$ fixed in the $k\rightarrow \infty$ limit
with $v\sim \sqrt{k}$. In this limit the deconstructed radii are
the same and correspond to the same spatial directions:
\begin{align}
&\lim_{k\rightarrow\infty} \frac{k}{g(v+\frac{\Delta}{2})} =
\lim_{k\rightarrow\infty} \frac{k}{g(v-\frac{\Delta}{2})} = R
\,, \nonumber \\
&\lim_{k\rightarrow\infty} \frac{g}{v+\frac{\Delta}{2}} =
\lim_{k\rightarrow\infty} \frac{g}{v-\frac{\Delta}{2}} = R_D \,.
\end{align}
At the point in moduli space given in (\ref{sep}), the quadratic
part of the effective $(2,0)$ superpotential is
\begin{align}
  W &= gv \int d\theta^+ {\rm tr} \left.\left[ B_j({\Lambda^{\tilde
          B}_-}_{j,j+1} - {\Lambda^{\tilde B}_-}_{j-1,j}) +
      {\Lambda^B_-}_{j,j+1}(\tilde B_{j+1} - \tilde
B_j)\right]\right\vert_{\bar \theta^+} \nonumber \\
  &+ \,g\Delta\int d\theta^+ {\rm tr} \left.\left[B_j({\Lambda^{\tilde
          B}_-}_{j,j+1} + {\Lambda^{\tilde B}_-}_{j-1,j}) +
      {\Lambda^B_-}_{j,j+1}(\tilde B_{j+1} + \tilde
      B_j)\right]\right\vert_{\bar \theta^+} \label{massterm} \,.
\end{align}
The second term in (\ref{massterm}) is a mass term from the point
of view of the lattice theory.  For large $k$ and fixed $n$,
diagonalizing the mass matrix for the fundamental spoke degrees of
freedom gives
\begin{align}
M^2 = (g\Delta)^2 + (n/R)^2\,,
\end{align}
where the integer $n$ is the lattice momentum obtained by Fourier
transforming with respect to the index $j$ labeling points on the
quiver. For simplicity let us set $n=0$, so that $m = g\Delta$. The
S-dual modes then have $m_D=\frac{k}{g}\Delta$. Since $m/m_D =
R_D/R$, the fundamental spoke degrees of freedom should be
interpreted as strings wrapping the cycle of radius $R_D$,  while
their S-duals wrap the cycle of radius $R$ (see figure 4). The
string tension is
\begin{align}
T = \frac{m}{2\pi R_D} = \frac{m_D}{2\pi R} = v\Delta \,.
\end{align}

\begin{figure}[!ht]
\begin{center}
\includegraphics{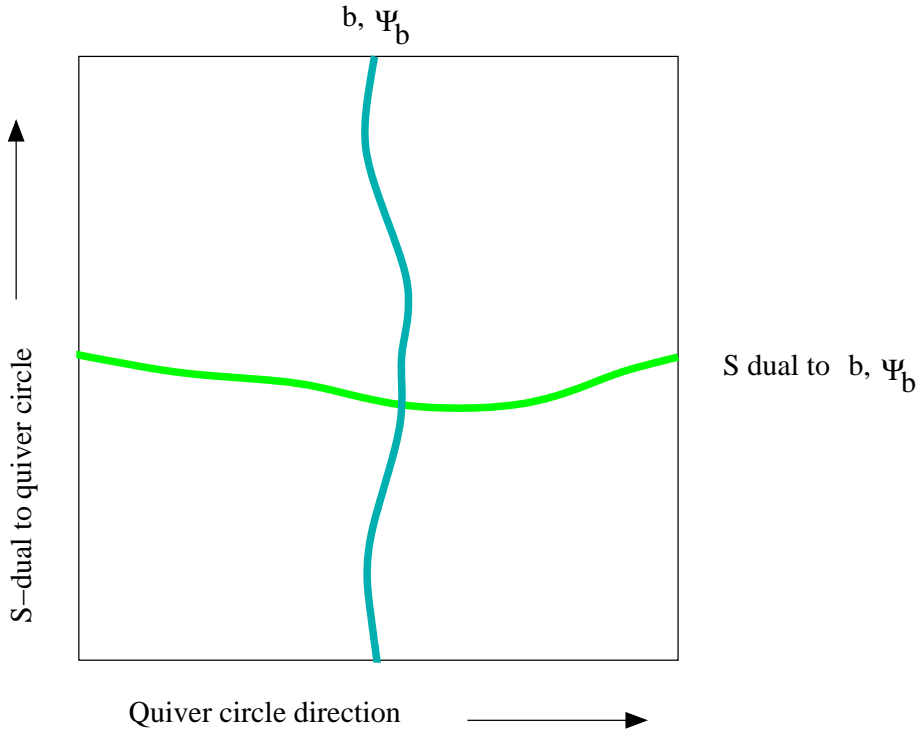}
\caption{Strings localized at the
intersection.}\label{wrappedstrings}
\end{center}
\end{figure}

Note that the S-dual's to the fundamental degrees of freedom at
the intersection are strings wrapping the $S^1$ of the quiver
diagram. Thus it is tempting to speculate that they can built from
gauge invariant products of fundamental spoke degrees of freedom
which wrap the quiver.  An example of such an operator is ${\rm
tr\,} \Lambda^B_{1,2}\Lambda^{\tilde B}_{2,3} \cdots
\Lambda^B_{k-1,k}\Lambda^{\tilde B}_{k,1}$. On the other hand, one
expects the S-dual operators to be solitons without an expression
in terms of products of local operators, so this speculation is
probably not quite correct.

\subsection{String condensation and M5-branes on a holomorphic curve}

When tensionless strings condense,  the M5-M5 intersection is
resolved to the holomorphic curve $xy =c$.  This can be seen very
explicitly from compactification on a torus.  In this case the low
energy theory is that of the D3-D3 intersection in flat space. In
section \ref{flat} we showed that the Higgs branch of the
corresponding $(4,4)$ dCFT can be interpreted as a resolution of
the intersection to the holomorphic curve $xy=c$. The resolved
intersection is also clearly captured by the $(4,0)$ dCFT.  At the
point in the moduli space for which extra dimensions are
generated,  the $(4,0)$ dCFT reduces to the $(4,4)$ dCFT at low
energies.  The potential is minimized by restricting to fields
with values independent of the quiver index $j$ and  satisfying
equations equivalent to (\ref{holomeq}).  The holomorphic curve
$xy=c$ arises when the fields $B_j$ and $\tilde B_j$ get
expectation values independent of $j$.  These fields correspond to
tensionless strings at the M5-M5 intersection.

\subsection{Identifying R-symmetries and moduli}

The M5-M5 intersection has ${\cal N} =2, d=4$ supersymmetry with
$SU(2) \times U(1)$ R-symmetry.  We  would like to identify the
corresponding charges in the $(4,0)$ defect conformal field
theory.

For the $U(1)$ R-symmetry,  the identification is as follows. This
symmetry is manifest in both cases and corresponds to a
simultaneous rotation of the $x$ and $y$ planes, which are
transverse to one stack of parallel branes but not the orthogonal
stack.  In the $(4,0)$ dCFT, it is generated by $J_{23}-J_{45}$,
and the associated charges can be readily obtained from table 1.
Note that the other linear combination, $J_{23} + J_{45}$, is not
an R-symmetry, and acts trivially on the degrees of freedom
localized at the intersection.

We will now argue that the $SU(2)_L$ R-symmetry of the $(4,0)$
dCFT should be identified with the $SU(2)$ R-symmetry of the
${\cal N} =2, d=4$ theory of the M5-brane intersection. This
matching is non trivial for the following reason. In order to
generate the extra dimensions, it was neccessary to consider a
point on the Higgs branch where the $SU(2)_L$ doublets $<Y>$ and
$<Y^{\prime}>$ are non-zero,  so that $SU(2)_L$ is spontaneously
broken.  On the other hand the $SU(2)$ R-symmetry of the M5-M5
intersection is only broken when M5-branes are transversely
separated. Nevertheless, we shall find evidence that the
identification makes sense.  This suggests that when the M5-branes
are not separated,  the $SU(2)_L$ symmetry of the $(4,0)$ dCFT
description is unbroken as far as the non-trivial dynamics is
concerned.

There are three directions transverse to both stacks of M5-branes,
corresponding to the moduli $\vec X$ and $\vec X^{\prime}$,  which
form a triplet under the $SU(2)$ R-symmetry.  This R-symmetry is
spontaneously broken if either $\vec X$ or $\vec X^{\prime}$ is
non-zero. However if all the eigenvalues of $\vec X$ and $\vec
X^{\prime}$ are the same,  then the symmetry breaking is due only
to trivial center of mass dynamics.  The string tension $T = |\vec
X- \vec X^{\prime}|$ vanishes at this point.

In the $(4,0)$ dCFT, the point in moduli space described by
(\ref{sep}) corresponds to a string tension $T= v\Delta$. If we
act with $SU(2)_L$,  we obtain another point in moduli space which
also deconstructs the same configuration of intersecting
M5-branes. The string tension can be written in an $SU(2)_L$
invariant way as the expectation value of $|Y^{\dagger}\vec\sigma
Y - {Y'}^\dagger \vec\sigma Y'|$ where $\vec \sigma$ are Pauli
matrices. We have dropped the $j,j+1$ subscript as we only
consider the zero momentum modes in the (de)constructed
directions.  On the other hand the string tension is related to
the moduli of the M5-M5 intersection by $T = |\vec X - \vec
X^{\prime}|$.  This motivates the proposal
\begin{align}
\vec X - \vec X^{\prime} \sim Y^{\dagger}\vec \sigma Y -
{Y'}^\dagger \vec\sigma Y' \,. \label{modreln}\end{align} Under
$SU(2)_L$, $Y^{\dagger}\vec \sigma Y- {Y'}^\dagger \vec\sigma Y'$
transforms as a triplet,  while $\vec X - \vec X^{\prime}$
transforms as triplet under $SU(2)$. This suggests that one should
identify the $SU(2)_L$ R-symmetry of the $(4,0)$ theory with the
$SU(2)$ R-symmetry of the M5-M5 intersection.

Thus far we have neglected a degree of freedom in the moduli space
which also contributes to the string tension. There are four
degrees of freedom in either $Y$ and or $Y'$, while only three are
characterized by $Y^{\dagger} \vec \sigma Y$ or ${Y'}^{\dagger}
\vec \sigma Y'$. Note that $Y^{\dagger} \vec \sigma Y$ is
invariant under $Y\rightarrow e^{i\theta} Y$, so the missing
degree of freedom is an angle.  The quantity
$|Y^{\dagger}\vec\sigma Y - {Y'}^\dagger \vec\sigma Y'| =
4vRe(\Delta)$ only gives the string tension for real $\Delta$. For
complex $\Delta$ the string tension is easily seen to be
$v|\Delta|$.  The imaginary part of $\Delta$ corresponds to the
additional angular degree of freedom. That the imaginary part is
an angle is evident from the orbifold condition $(u,\bar w) \sim
\exp(2\pi i/k) (u,\bar w)$ which for large $k$ gives $\Delta \sim
\Delta + 2\pi i v/k$.  By viewing the quiver action as an action
with only one extra discretized dimension (i.e. taking $R_D
\rightarrow 0$ with fixed $R$), one discovers that the angular
degree of freedom is a gauge connection in the compact discretized
fifth direction. If the associated Wilson lines differ for the two
intersecting branes, a mass term is generated for the degrees of
freedom localized at the intersection. In terms of the six
dimensional theories, this Wilson line may be interpreted as a
Wilson surface corresponding to the holonomy of the mysterious
non-abelian two-form on the torus.

\subsection{A 't Hooft anomaly}

Although the deconstruction of the M5-M5 intersection involves a
strong coupling limit,  certain protected quantities such as 't
Hooft anomalies may be computed.   We shall find a 't~Hooft
anomaly in the $SU(2)_L$ R-current of the $(4,0)$ defect CFT.
This means that there are Schwinger terms for correlators of the
$SU(2)_L$ current which imply that the current is not conserved
upon coupling to a background $SU(2)_L$ gauge field.  For the
M5-M5 intersection, this result suggests a 't Hooft anomaly in the
$SU(2)$ R-symmetry of the M5-M5 intersection due to tensionless
strings.

Before discussing this anomaly, let us revisit the R-symmetry
't~Hooft anomaly which is known to arise for coincident M5-branes.
The 't~Hooft anomaly in the $Spin(5)$ R-symmetry of the
six-dimensional $(2,0)$ theory has been derived from anomaly
cancellation considerations in M-theory \cite{HMM}, but never
directly from a microscopic formulation of the $(2,0)$ theory,
except in the abelian case \cite{Witteneff}. The $Spin(5)$
R-symmetry can be viewed as part of the unbroken Lorentz symmetry
for M-theory in the presence of flat parallel 5-branes. Since
M-theory includes eleven-dimensional supergravity, this R-symmetry
is actually gauged.  The basic idea of \cite{HMM} was to consider
the long wavelength components of eleven-dimensional supergravity
in the presence of a magnetic source, i.e. five-branes.  The
relevant terms in the supergravity action are
\begin{align}
\int_{M_{11}} \, \left(C_3 \wedge I_{8}^{inf} - \frac{2\pi}{6} C_3
\wedge  G_4 \wedge G_4\right) \,, \label{cernsimons}
\end{align}
where $G_4 = d C_3$. In the presence of a magnetic source, such
that
\begin{align}
d G_4 = Q_5 \Pi_{i=6}^{10} \delta(x^i) dx^0 \wedge dx^1 \cdots
\wedge dx^5\, , \label{source}
\end{align}
the term (\ref{cernsimons}) is not invariant under diffeomorphisms
of the normal bundle.\footnote{To see that the second term in
(\ref{cernsimons}) is not diffeomorphism invariant is subtle, and
requires a regulation of the delta function in (\ref{source})
\cite{FHMM,HMM}.} Assuming that this anomaly is cancelled requires
that the degrees of freedom on the M5-branes produce the opposite
anomaly. In the decoupling limit,  this anomaly becomes a 't Hooft
anomaly (or Schwinger term) for the global $Spin(5)$ R-symmetry of
the $(2,0)$ theory.

Attempts to calculate this anomaly in the six-dimensional $(2,0)$
theory using its (de)con\-structed description \cite{Arkani-Hamed}
quickly run into a difficulty.  The difficulty arises because the
$Spin(5)$ R-symmetry of the $(2,0)$ theory is not manifest in the
deconstructed description,  which is a ${\cal N} =2, d=4$
superconformal Yang-Mills theory having only a $SU(2) \times U(1)$
R-symmetry.  The full $Spin(5)$ symmetry can only arise in the
continuum limit.  Since the $SU(2) \times U(1)$ R-symmetry of the
four-dimensional theory does not exhibit a 't Hooft
anomaly\footnote{There is however a global $SU(2)$ anomaly if $k$
is odd \cite{WittenSU(2)}.}, one can not obtain information about
the $Spin(5)$ anomaly without a detailed understanding of the
enhancement of $SU(2) \times U(1)$ to $Spin(5)$ in the continuum
limit.  Note that even if it is possible to gauge a subgroup of a
global symmetry without encountering an anomaly, the same may not
be true for the full symmetry group.

In the case of intersecting M5-branes,  we shall find that there
is an additional anomaly which follows directly from the
description in terms of the $(4,0)$ gauge theory. The $SU(2)$
R-symmetry of the intersecting five-branes corresponds to the
$SU(2)_L$ R-symmetry of the $(4,0)$ gauge theory description,
which exhibits a 't Hooft anomaly. In a four-dimensional field
theory, $SU(2)$ is free of 't Hooft anomalies. Thus a contribution
to a $SU(2)$ anomaly can only come from the two-dimensional
degrees of freedom in the $(4,0)$ gauge theory. The only
two-dimensional fermions charged under $SU(2)_L$ are contained in
the $(4,0)$ hypermultiplets and have positive chirality. These
make up a set of $k$ positive chirality doublets $(\Psi^+_{b,j},
\bar \Psi^+_{\tilde b,j})$ in the representation $(N,\bar
N^{\prime})$ of $SU(N)_j \times SU(N^{\prime})_j$.  Thus upon
introducing a background $SU(2)_L$ gauge connection $\tilde A$,
one finds the anomaly
\begin{align}
D_m \frac{\delta S_{eff}(\tilde A)}{\delta \tilde A_m^a(z)} =
\frac{g^2}{4\pi}kNN^{\prime}\epsilon_{mn}\tilde F^{mn\, a}(z) \,,
\label{anomaly1}
\end{align}
where $m,n=0,1$.  Some brief clarifying comments are in order
about the meaning of the background $SU(2)_L$ gauge connection
appearing in (\ref{anomaly1}).  This connection can be regarded as
two connections on each of the intersecting world volumes of the
dCFT, ${\cal A}_{\mu}(z,x)$ for $\mu = 0,1,2,3$ and ${ \cal
A}^{\prime}_{\alpha}(z,y)$ for $\alpha = 0,1,4,5$, subject to the
constraint ${\cal A}_{0,1}(z,0) = {\cal A}^{\prime}_{0,1}(z,0)
\equiv \tilde A_{0,1}(z)$.  In terms of a current equation,  the
anomaly for the dCFT coupled to an $SU(2)_L$ gauge connection is
\begin{align}
\delta^2(y)D_{\mu}J^{\mu}(z,x) + \delta^2(x)D_{\alpha}J^{\prime
\alpha}(z,y) + \delta^2(x)\delta^2(y) &D_m {\cal J}^{m}(z) =
\nonumber \\
&\delta^2(x)\delta^2(y)\frac{g}{4\pi}kNN^{\prime}\epsilon^{mn}{\tilde
F}_{mn}(z) \, ,
\end{align}
where $J$, $J^{\prime}$ and ${\cal J}$ depend on fields living on
the D3-brane, the D3$^\prime$ brane, and the intersection
respectively.

The question is now whether the anomaly (\ref{anomaly1})
corresponds to a {\it finite} anomaly of the $SU(2)$ R-symmetry of
the M5-M5 intersection in the $k\rightarrow\infty$ limit. If
finite, the continuum limit should be interpreted as an anomaly
arising from tensionless strings propagating in four dimensions.
Note that while a local $SU(2)$ anomaly is not possible for a
four-dimensional field theory,  an $SU(2)$ anomaly of the M5-M5
intersection would be due to tensionless strings rather than local
quantum fields! Unfortunately,  we do not yet know how to obtain
the continuum limit of (\ref{anomaly1}),  which can presumably be
viewed as a sum of a discretized four-dimensional anomaly equation
over the lattice sites.

An alternate way to derive the anomaly is to look for
diffeomorphism anomalies of eleven-dimensional supergravity in the
presence of magnetic sources due to intersecting M5-branes. The
$SU(2)$ R-symmetry of the M5-M5 intersection is the Lorentz
symmetry rotating the directions transverse to the intersecting
five-branes. Thus in eleven-dimensional supergravity, the $SU(2)$
't Hooft anomaly becomes an anomaly in diffeomorphisms of the
normal bundle. This should presumably be cancelled by an anomaly
due to long-wavelength terms of the supergravity action in the
presence of magnetic sources.  We have not as yet been able to
show this, however we will comment briefly on the contribution of
the Chern-Simons terms (\ref{cernsimons}) to the anomaly.

The first term in (\ref{cernsimons}) is linear in $C^3$, while the
second term is cubic.  Only the cubic term can contribute to an
anomaly localized at the five-brane intersection, due to mixed
terms arising from the magnetic source
\begin{align}
d G_4 = &N\Pi_{i=6,7,8,9,10} \, \delta(x^i)\, dx^0 \wedge dx^1 \wedge
dx^2 \wedge dx^3 \wedge dx^4 \wedge dx^5 \nonumber \\ +
&N^{\prime}\Pi_{i= 4,5,8,9,10}\, \delta(x^i)\,  dx^0 \wedge dx^1
\wedge dx^2 \wedge dx^3 \wedge dx^6 \wedge dx^7\, .
\end{align}
If the Chern-Simons term gives a non-zero anomaly, the coefficient
will be proportional to $(N+N^{\prime})NN^{\prime}$. This
apparently does not match the $N$ and $N^{\prime}$ dependence of
(\ref{anomaly1}). Perhaps the correct dependence somehow arises
from correctly lifting (\ref{anomaly1}) to a continuous
four-dimensional version.

\section{Conclusion and open questions}

In this paper we have presented a formulation of intersecting
M5-branes in terms of a limit of a $(4,0)$ defect conformal field
theory.  We hope that this will lead to an improved understanding
of the low energy dynamics of M5-branes although, as for the
(de)construction of parallel M5-branes \cite{Arkani-Hamed},
immediate progress is impeded by the fact that the continuum limit
is also a strong coupling limit.

At the moment we only have control of a few some simple properties
which are protected against radiative and non-perturbative
corrections, such as the 't Hooft anomaly in the $SU(2)$ R-current
due to tensionless strings.  This anomaly clearly deserves further
study. In particular the contribution of M-theory Chern-Simons
terms to the anomaly should be computed.

It would be interesting to try to generalize the construction here
to more complicated intersections of branes in M-theory.  Such
generalizations might be of use in understanding the microscopic
origins of black hole entropy.

It would also be very interesting to find field theoretic
arguments in favor of the S-duality of the D3-D3 intersection,
either in flat space or at a $\CC^2/\ZZ_k$ orbifold.  We have only
assumed S-duality in this paper,  based on the S-duality of the
string theory background.  A starting point would be to find
solitons which are S-duals of degrees of freedom localized at the
intersection. This is clearly very important if one wishes to have
a better understanding of the degrees of freedom and dynamics of
intersecting M5-branes.

\vspace{2cm}
\section*{Acknowledgements}

The authors wish to thank G. Cardoso, A. Hanany, R. Helling, B.
Ovrut, S. Ramgoolam, W.~Skiba, D.~Tong and J.~Troost for helpful discussions.
The research of J.E., Z.G.~and I.K.~is
funded by the DFG (Deutsche Forschungsgemeinschaft) within the
Emmy Noether programme, grant ER301/1-2. N.R.C.\ is supported by
the DOE under grant DF-FC02-94ER40818, the NSF under grant
PHY-0096515 and NSERC of Canada.

\newpage
\appendix

{\noindent \Large \bf Appendix}

\vspace{1cm}

\section{Gauge transformation properties}  \label{transformations}
\setcounter{equation}{0}

The gauge transformation properties under the residual gauge group
$SU(N)^k\times SU(N')^k$ in $(2,0)$ superspace are as follows:\
\begin{align} \label{transform}
  \tilde B_i &\rightarrow e^{-i \lambda'_i} \tilde B_i e^{i \lambda_i}, \quad
  \Lambda^{\tilde B}_{-,i} \rightarrow e^{-i \lambda'_{i}}
  \Lambda^{\tilde B}_{-,i} e^{i\lambda_{i-1}}, \nonumber \\
  B_i &\rightarrow e^{-i \lambda_i} B_i e^{i \lambda'_{i}},\quad
  \Lambda^{B}_{-,i} \rightarrow e^{-i \lambda_{i}} \Lambda^{B}_{-,i}
  e^{i \lambda'_{i-1}}   \,, \nonumber\\
  Q^1_i &\rightarrow e^{-i \lambda_i} Q^1_i e^{i \lambda_{i+1}},\quad
  \Lambda_{-,i}^{Q^1} \rightarrow e^{-i \lambda_i} \Lambda_{-,i}^{Q^1}
  e^{i \lambda_i} , \\
  Q^2_i &\rightarrow e^{-i \lambda_i} Q^2_i e^{i \lambda_{i}},\quad
  \Lambda_{-,i}^{Q^2} \rightarrow e^{-i \lambda_i} \Lambda_{-,i}^{Q^2} e^{i
    \lambda_{i+1}} ,
  \nonumber\\
  \Theta^V_i &\rightarrow e^{-i \lambda_{i}} \Theta^V_i e^{i \lambda_{i-1}},
  \quad
  e^{V_i} \rightarrow e^{-i \lambda^\dagger_i} e^{V_i} e^{i \lambda_i},  \nonumber\\
  S^1_i &\rightarrow e^{-i \lambda'_i} S^1_i e^{i \lambda'_{i+1}},\quad
  \Lambda_{-,i}^{S^1} \rightarrow e^{-i \lambda'_i} \Lambda_{-,i}^{S^1} e^{i
    \lambda'_i} ,
  \nonumber\\
  S^2_i &\rightarrow e^{-i \lambda'_i} S^2_i e^{i \lambda'_{i}},\quad
  \Lambda_{-,i}^{S^2} \rightarrow e^{-i \lambda'_i} \Lambda_{-,i}^{S^2} e^{i
    \lambda'_{i+1}} ,
  \nonumber\\
  \Theta^\V_i &\rightarrow e^{-i \lambda'_{i}} \Theta^\V_i e^{i
    \lambda'_{i-1}}, \quad e^{{\cal V}_i} \rightarrow e^{-i
    {\lambda'}^\dagger_i} e^{{\cal V}_i} e^{i \lambda'_i} \nonumber\,,
\end{align}
where $i=1,\dots k$. These gauge transformations lead to the
quiver diagram of figure \ref{superquiver}. Strictly speaking, the
transformation laws for the Fermi multiplets hold only at $\bar \theta^+=0$.
Note that Fermi multiplets in superpotentials act effectively  as chiral
multiplets.

\section{Superpotential in manifest $(2,0)$ language}
\setcounter{equation}{0}

The conformal field theory corresponding to the D3-D3 intersection placed at
an orbifold singularity is $(4,0)$ supersymmetric. For an adequate
formulation
of the parent defect theory in flat space, we use $(2,0)$ superspace. For
the
purposes of this paper it is sufficient to give the superpotential of the
parent theory, which we now express in terms of $(2,0)$ superfields. Writing
the full $(4,4)$ supersymmetric action (\ref{bulkaction1}) -
(\ref{fullimpaction}) in $(2,0)$ superspace is straightforward, but we do
not
give the result here.

We decompose the $(2,2)$ defect multiplets $\bf B$ and
$\bf \tilde B$ as well as the ambient multiplets $\bf Q_1$, $\bf Q_2$, $\bf
\Phi$, and $\bf V$ under
(2,0) supersymmetry. The reduction of $(2,2)$ multiplets to $(2,0)$
multiplets is
discussed in \cite{Witten, Garcia}.

In general, a $(2,2)$ chiral multiplet $\bf \Phi$ reduces to two
$(2,0)$ multiplets, a chiral multiplet $\Phi$ and a
Fermi multiplet $\Lambda_-$, according to
\begin{align} \label{chiralexp}
  \mathbf{\Phi}(y,\theta^\pm)=\Phi(y,\theta^+,\bar\theta^+)
  \vert_{\bar\theta^+=0} + \sqrt{2} \theta^-
\Lambda_-(y,\theta^+,\bar\theta^+)
  \vert_{\bar\theta^+=0} \,
\end{align}
with $y^M=x^M + i \theta^+ \bar \theta^+ + (-1)^M i \theta^- \bar \theta^-$
($M=0,1$).  The chiral $(2,0)$ multiplet satisfies ${\cal \bar D}_+
\Phi= 0$ and can be expanded as
\begin{align}
\Phi = \phi+\sqrt{2}\theta^+ \lambda_+
- i \theta^+ \bar \theta^+ (D_0 + D_1) \phi \,
\end{align}
with covariant derivatives $D_M=\partial_M +\frac{ig}{2} v_M$.
The Fermi multiplet expansion is given by
\begin{align}
  \Lambda_- = \psi_- + \sqrt{2} \theta^+ F
  -i \theta^+ \bar \theta^+ (D_0 + D_1) \psi_- - \sqrt{2} \bar \theta^+ E
\,.
\end{align}
$\Lambda_-$ satisfies ${\cal \bar D}_+ \Lambda_- = \sqrt{2} E$.  In the reduction
of the above $(2,2)$ chiral superfield $\bf \Phi$, the function $E$ is $E= i
\sqrt{2} T^a
\Theta^a_V \Phi$, where $T_a$ are the generators of the gauge group. Here
$\Theta_V$ is another chiral superfield defined by
\begin{align}
\Theta_V \equiv \mathbf{\Sigma}
  \vert_{\theta^-=\bar\theta^-=0}=\s + i \theta^+ \bar \lambda_+
  - i \theta^+ \bar \theta^+ (D_0 + D_1) \s \,,
\end{align}
where $\bf \Sigma$ is the gauge invariant field strength of the
(2,2) gauge multiplet $\bf V$.

We can now write the superpotential $W^{\rm par}
=W^{\rm par}_{\rm D3} + W^{\rm par}_{\rm D3} +W^{\rm par}_{\rm D3-D3'}$
of the parent theory, by substituting the following expansions
into the action (\ref{bulkaction1})-(\ref{fullimpaction}),
\begin{align}
  \mathbf{Q}_i&= (Q_i+\sqrt{2}\theta^- \Lambda^{Q_i}_-
  )\vert_{\bar\theta^+=0} \,,\qquad
  \mathbf{B}=(B+\sqrt{2}\theta^- \Lambda^{B}_-)\vert_{\bar\theta^+=0}
  \,, \\
  \mathbf{\Phi}&= (\Phi+\sqrt{2}\theta^- \Lambda^{\Phi}_-
  )\vert_{\bar\theta^+=0} \,,\qquad\,\,\,\,
  \mathbf{\tilde B}=(\tilde B +\sqrt{2}\theta^- \Lambda^{\tilde B}_-)
  \vert_{\bar\theta^+=0} \,.
\end{align}
For the superpotential $W^{\rm par}_{\rm D3}$ associated with one stack
of D3-branes, we
find
\begin{align}
  W^{\rm par}_{\rm D3} = & \int d^2x d^2 \theta\, \epsilon_{ij} {\rm\, tr\,}
  \mathbf{Q}_i
  [\partial_{\bar x} + g \mathbf{\Phi}, \mathbf{Q}_j] + c.c \\
  =& \int d^2x d \theta^+ {\rm tr\!}\left.\left( \Lambda_-^{Q_1} [
      \partial_{\bar x} + g \Phi, Q_2] -\Lambda_-^{Q_2} [ \partial_{\bar x}
+
      g \Phi, Q_1] + g \Lambda_-^{\Phi} [Q_2, Q_1] \right)
  \right\vert_{\bar\theta^+=0} \!+\! c.c.  \nonumber
\end{align}
A similar expression holds for $W^{\rm par}_{\rm D3'}$, while the defect
action has the $(2,0)$ superpotential
\begin{align} \label{sp}
  W^{\rm par}_{\rm D3-D3'}= \frac{ig}{2}\int d\theta^+ {\rm\, tr} &\left(
      B \tilde B \Lambda_-^{Q_1} + \Lambda^{B}_- \tilde B Q_1 + B
      \Lambda^{\tilde B}_- Q_1 \right. \nonumber\\
&-  \tilde B B \Lambda_-^{S_1} - \Lambda^{\tilde B}_- B S_1 - \tilde B
      \Lambda^{B}_- S_1\left.
\big) \right\vert_{\bar \theta^+=0} +c.c.
\end{align}
The parent superpotential $W^{\rm par}$ leads to the superpotential
(\ref{super}) under the orbifold projection.

\end{document}